%% file: ntorrent-ndnsim-port.tex
\documentclass[10pt, conference, letterpaper]{IEEEtran}

\usepackage{hyperref}
\usepackage{array}
\usepackage{graphicx}
\usepackage[table]{xcolor}
\usepackage{algpseudocode}
\usepackage{hyperref}

\newcolumntype{C}[1]{>{\centering\let\newline\\\arraybackslash\hspace{0pt}}m{#1}}
\algnewcommand\algorithmicinput{\textbf{Input:}}
\algnewcommand\INPUT{\item[\algorithmicinput]}



\title{Porting nTorrent to ndnSIM}
\author{
\IEEEauthorblockN{Kimberly Chou,  Akshay Raman}
\IEEEauthorblockA{University of California, Los Angeles \\  \{klchou,  akshay.raman\}@cs.ucla.edu}
}

\usepackage{etoolbox}
\usepackage{xstring}
\DeclareListParser{\doslashlist}{/}
\newcounter{ndnNameComponentCounter}%
\newcommand{\ndnName}[1]{{%
  \setcounter{ndnNameComponentCounter}{0}%
  \renewcommand{\do}[1]{{%
    \ifnumgreater{\value{ndnNameComponentCounter}}{0}{\allowbreak/}{}%
    \ifnumodd{\value{ndnNameComponentCounter}}{}{}%
    \detokenize{##1}}%
    \stepcounter{ndnNameComponentCounter}}%
``{\fontfamily{cmtt}\small\selectfont\IfBeginWith{#1}{/}{/}{}\doslashlist{#1}}''%
}}

\usepackage{eso-pic,xcolor}
\makeatletter
\AddToShipoutPicture*{%
\setlength{\@tempdimb}{20pt}%
\setlength{\@tempdimc}{\paperheight}%
\setlength{\unitlength}{1pt}%
}
\makeatother

\usepackage{color,soul}
\usepackage{xspace}

\begin{document}
\maketitle

\begin{abstract}

BitTorrent is a popular communication protocol for peer-to-peer file sharing. It uses a data-centric approach, wherein the data is decentralized and peers request each other for pieces of the file(s). Aspects of this process is similar to the Named Data Networking (NDN) architecture, but is realized completely at the application level on top of TCP/IP networking. nTorrent is a peer-to-peer file sharing application that is based on NDN. The goal of this project is to port the application onto ndnSIM to allow for simulation and testing.

\end{abstract}

\input{intro}
\input{related-work}
\input{design}
\input{simulations}
\input{conclusion}

\section*{Acknowledgments}

We would like to thank Spyridon Mastorakis for providing us with all the help needed to pursue this project.

\bibliographystyle{plain}
\bibliography{refs}

\end{document}

%% file: intro.tex
\section{Introduction}

Named Data Networking (NDN)~\cite{zhang2010named} is a network layer protocol that is being actively researched with the hope of serving as a replacement for the IP protocol. nTorrent~\cite{mastorakis2017ntorrent} is an NDN peer-to-peer file sharing application. The current implementation runs with a few modifications to the base ns-3 network simulator in order to compile and run successfully. The idea behind this paper is to extend the functionality of nTorrent and make it run on top of ndnSIM~\cite{mastorakis2017evolution, mastorakis2016ndnsim, mastorakis2015ndnsim} that features full integration with the NDN Forwarding Daemon (NFD)~\cite{nfd-dev} for simulations.

Our code is available at \url{https://github.com/akshayraman/scenario-ntorrent}.

%% file: related-work.tex
\begin{figure}[h]
  \centering
  \includegraphics[width=\columnwidth]{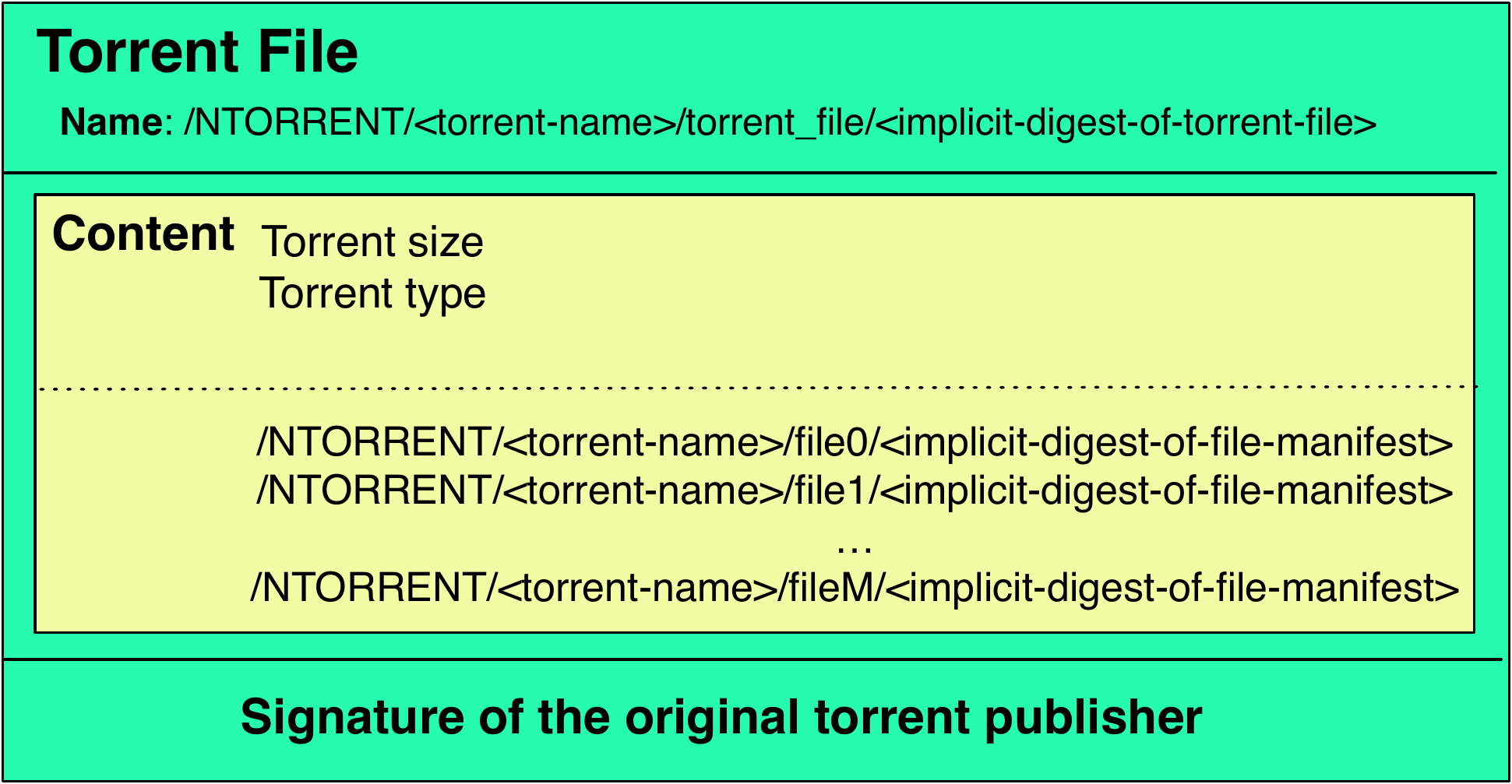}
  \caption{\small Structure of a torrent-file}
  \label{Figure:torrent-file}
\end{figure}
\begin{figure}[h]
  \centering
  \includegraphics[width=\columnwidth]{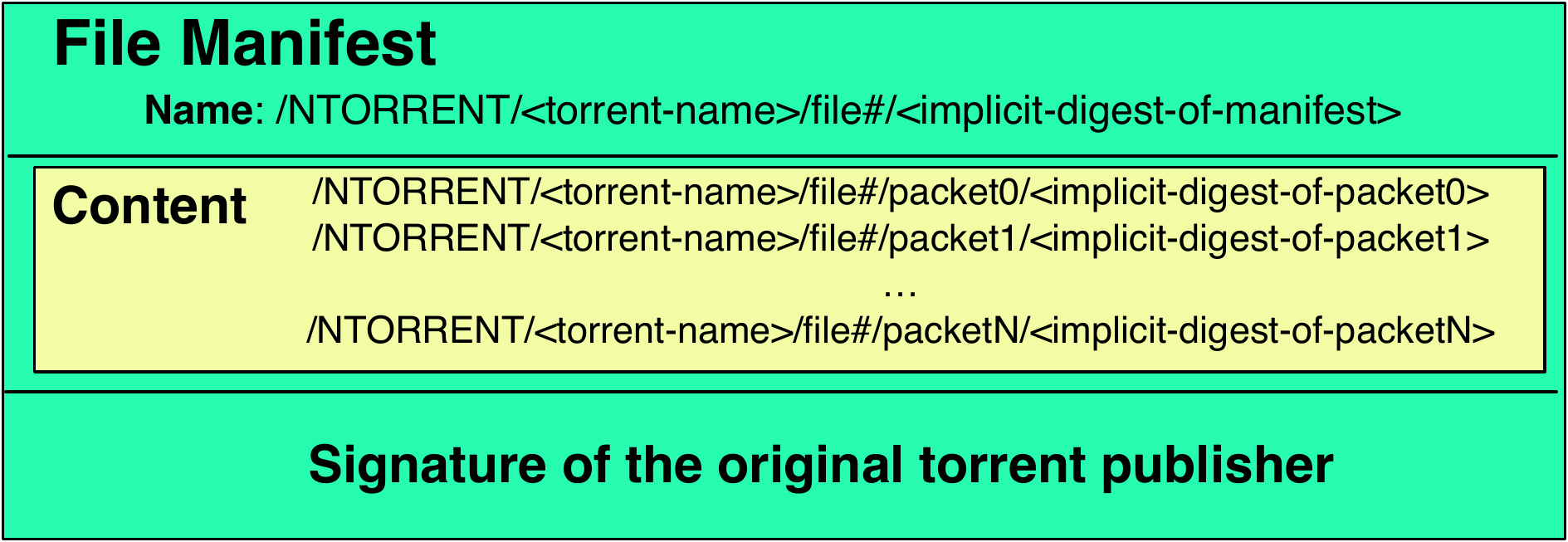}
  \caption{\small Structure of a file manifest}
  \label{Figure:file-manifest}
\end{figure}

\section{Related Work}

nTorrent has been designed to have a hierarchical file structure. At the top of the hierarchy, the .torrent file (Figure~\ref{Figure:torrent-file}) contains the name, size, type of the torrent file, and the signature of the original publisher. It also includes the names of the file manifests that make up the torrent file. Each file manifest (Figure~\ref{Figure:file-manifest}) contains its name, signature of the original publisher, and a list of names of the packets that make up that particular file. Using these names, Interest packets can be sent out to request for the corresponding files or packets. 

The fetching strategy currently implemented starts with requesting for the first packet of the first file to the last packet of the last file. Each name also follows a certain naming convention that can help easily identify the name of the torrent file, the file names, and the individual packets in a file. This name is also used by the routers to verify the integrity of the the packet.

%% file: design.tex
\section {Implementation Design}

In this section, we describe our high-level implementation design logic and elaborate on implementation-specific details.

\subsection {Implementation Design Logic}

ndnSIM currently supports the simulation of: (i) applications written against the ndn-cxx library (called \emph{real-world} applications) ported to ndnSIM, such as RoundSync~\cite{de2017design, roundsync-template} for decentralized data synchronization and ndn-ping~\cite{ndn-tools}, and (ii) ndnSIM-specific applications realized based on NS3's Application abstraction, such as those developed for experimentation with Fuzzy Interest Forwarding (FIF)~\cite{Chan:2017:FIF:3154970.3154975, mastorakis2018experimentation, fif-repo} and a best-effort scheme for link reliability~\cite{vusirikala2016hop, reliability-repo}. 

The existing nTorrent code~\cite{ntorrent-code} is an application written against the ndn-cxx library, but, for now, we have taken a hybrid approach between (i) and (ii). Specifically, we compile nTorrent as a shared library and import, and reuse its data structures (e.g., torrent-file, file manifest, etc.) into ndnSIM-specific applications. We are planning to move toward approach (i), as we further develop the codebase.

\subsection {Implementation Design Specifics}

Using the ndnSIM scenario template~\cite{scenario-template} and the ChronoSync simulation example~\cite{scenario-chronosync},  we have created two extensions - one for the consumer and the other for the producer. The example simulation in this project is called ?ntorrent-simple? and this is the entry point for the simulation. In this script, we configure the attributes of the system (data rate, latency and so on) as well as the individual attributes for the producer and consumer. These attributes include number of names per torrent segment, number of names per manifest and the size of each data packet. All these variables have default values but can be easily modified via the command-line.

The producer is responsible for generating the torrent file segments, manifests and data packets. The code to generate the above is being reused from the existing nTorrent implementation. All of this is stored in private variables belonging to the producer. The number of files being transferred is a constant that can be configured in the simulation-constants.hpp file. These are not real files - they are just pretend files to illustrate the working of nTorrent. 

The internal functionality of some modules in the nTorrent codebase was re-implemented to skip file I/O operations and use data structures (vectors) instead. For instance, all bytes read as a result of a file read operation were replaced with a dummy character "A" to simulate a file read. The producer extension also has functionality to implement the OnInterest method. This method is responsible for processing the interests that it receives from the consumer. In this method, we determine the type of the interest received. The interest could be of type torrent segment, file manifest or data packet. After determining the type of interest, the producer responds to these interests accordingly. If the interest is for a torrent file segment, the producer first checks if it has  the requested segment. If so, it responds with the segment. The file manifest and data packet interest-response functionality is also handled similarly. 

The consumer has a copyTorrentFile method to simulate the copy of the torrent file. In the BitTorrent model, the consumer and producer are both expected to have the torrent file. This is achieved in the simulation by generating the torrent file on the consumer end too. As the NDN communication model is consumer-driven, the consumer starts the file transfer with an interest for the first torrent segment, contained in this torrent file. Additionally, the consumer implements the OnData and SendInterest methods. The SendInterest method is used to send out an interest packet. This method appends a random number, the nonce, to prevent replay attacks. All interests are made using the nonce and the sha256digest. The OnData method implements the core functionality of the consumer. It is structurally similar to the OnInterest method in the producer. 

If the consumer receives a torrent file segment as the data, it uses the pointer to the next torrent segment and sends out an interest for this segment. It also iterates over the manifest catalog and sends interests for all the associated manifests. Similarly, if the consumer receives file manifest data, it uses the next manifest pointer to send an interest for the next file manifest. The associated sub-manifest catalog is iterated over to send interests for data packets. Finally, if the consumer receives a data packet, it is decrypted and displayed on the terminal. The consumer also stores all torrent segments, file manifests and data packets as private data members  in vectors in lieu of actual files. The simulation terminates once all data packets have been received. The success of the file transfer can be verified using the Content Store and interface statistics. 

%% file: simulations.tex
\section {Simulation Screenshots}

\begin{figure}[h]
  \centering
  \includegraphics[width=\columnwidth]{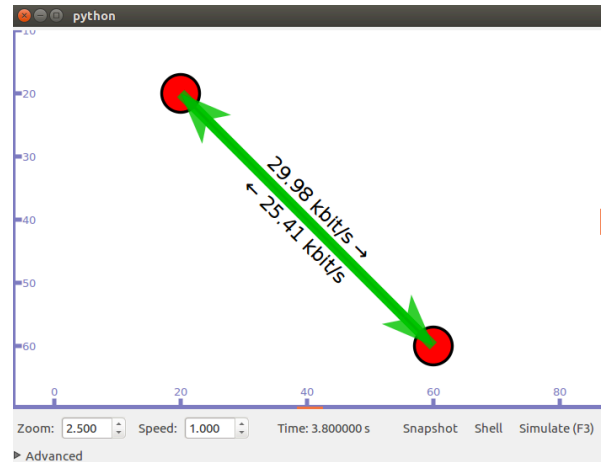}
  \caption{\small The simulation setup. The node in the top-left is the producer and the bottom-right node is the consumer.}
  \label{Figure:setup}
\end{figure}

In this section, we present a few simulation screenshot using the visualizer tool of ns-3. For the sake of simplicity, we assume a toy example topology consisting of a node that acts as a producer and a node that acts as a consumer (Figure~\ref{Figure:setup}). The consumer node first requests the torrent-file from the producer. For now, the implemented fetching strategy\footnote{A fetching strategy that maximizes the efficiency of the retrieval process is a part of our future work.} requires that a consumer fetches all the segments of the torrent-file before it starts downloading the file manifests. 

\begin{figure}[h]
  \centering
  \includegraphics[width=\columnwidth]{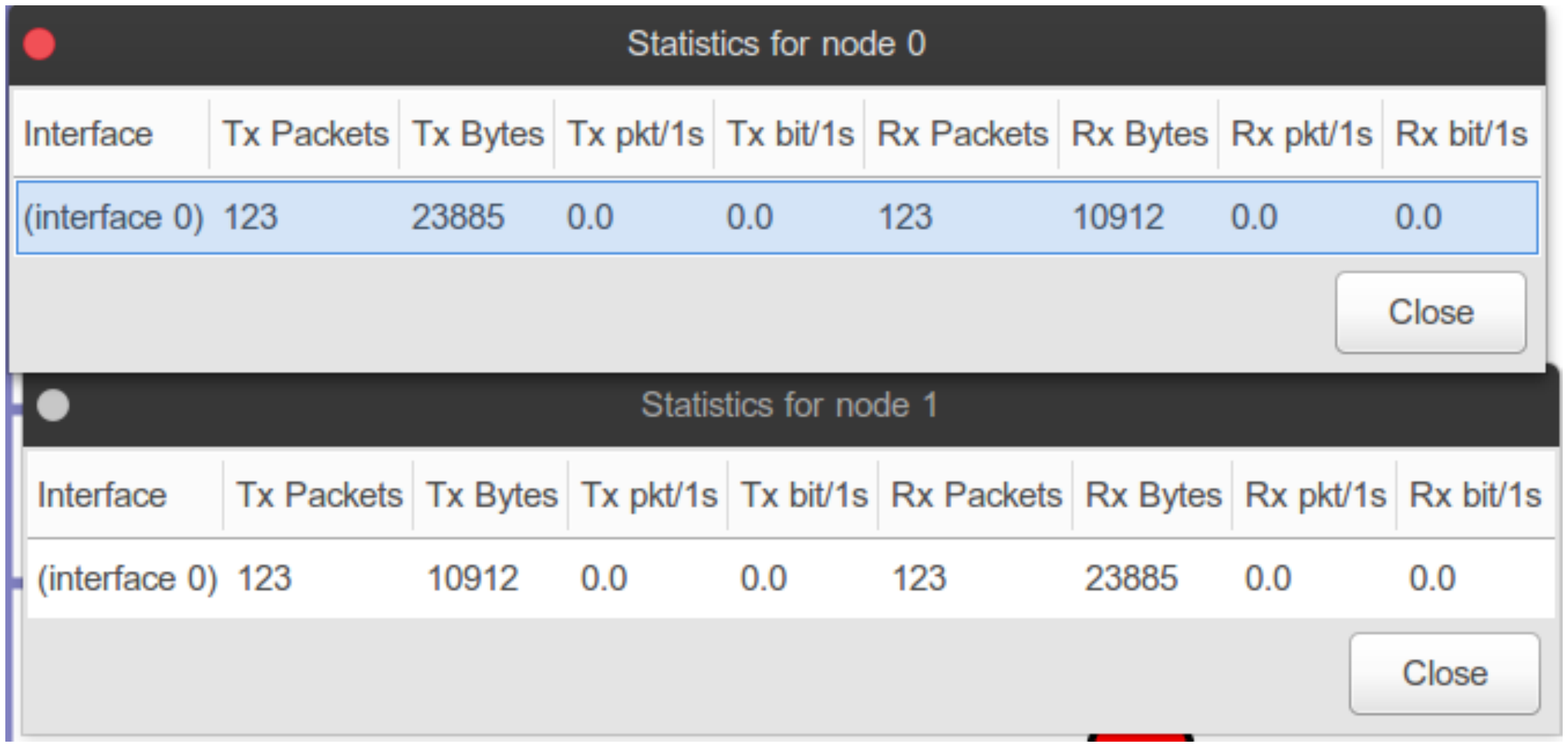}
  \caption{\small Data sent by the producer (node 0) and received by the consumer (node 1)}
  \label{Figure:node0-node1}
\end{figure}

\begin{figure}[h]
  \centering
  \includegraphics[width=\columnwidth]{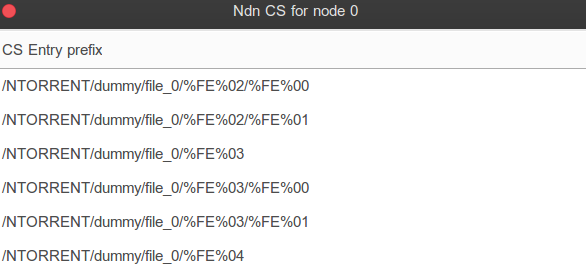}
  \caption{\small A view of the consumer's Content Store (CS)}
  \label{Figure:cs}
\end{figure}

\begin{figure}[h]
  \centering
  \includegraphics[width=\columnwidth]{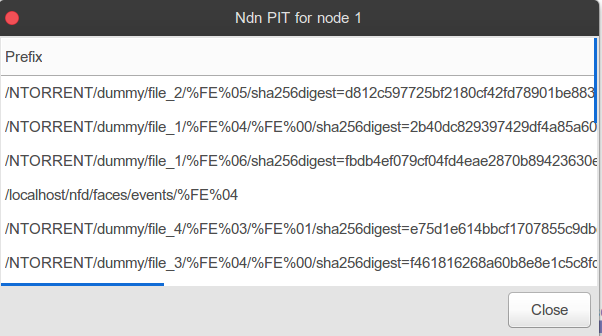}
  \caption{\small A view of the consumer's Pending Interest Table (PIT)}
  \label{Figure:pit}
\end{figure}

After all the file manifests have been fetched, the consumer starts downloading the individual data packets from the producer. Specifically, the consumer expresses Interests iteratively for all the data packets included in each file manifest. The overall process is illustrated in Figure~\ref{Figure:node0-node1} and screenshots of the consumer's CS and PIT are illustrated in Figure~\ref{Figure:cs} and~\ref{Figure:pit}.

%% file: conclusion.tex
\section {Conclusion and Future Work}

This  paper  describes  our  design  and  experimentation  with ndnSIM and nTorrent. This project is a clear example of how the NDN model is "consumer-driven". This project has started gaining enquiries on the ndnSIM mailing list despite being it still being a work in progress and just over a month old.

In the real world, an nTorrent node should be able to send and receive data from other nodes. That is, it should be able to act as both a producer and a consumer. We plan to support this bifunctional peer-to-peer model in the simulation in the future. To achieve that, we should be able to make routing announcements from a node when it has received the complete torrent file, file manifest or data packet. In the current implementation, we decided not to deal with file I/O. We plan to support real file handling and splitting in the future. Going forward, we would also like to support two modes of compilation: one as a standalone application and the other as an ndnSIM scenario. Furthermore, we want to investigate the scalability and trade-offs of this nTorrent design with more tests and simulations, as well as the fetching strategies that maximize the retrieval efficiency.